\documentclass[11pt]{article}
\usepackage{amsmath,amssymb,cite,dsfont,mathtools,tensor,color}
\usepackage{hyperref,verbatim,ulem}
\usepackage{graphics,float,overpic}
\usepackage{epsfig}
\usepackage{overpic} 
\usepackage{array,multirow}
\usepackage{enumerate}

\newcommand{\ket}[1]{\left|  #1\right\rangle}

\numberwithin{equation}{section}

\title{The matrix model of two-color one-flavor QCD: \\
The ultra-strong coupling regime}

\author{Nirmalendu Acharyya$^a$\footnote{nirmalendu@iitbbs.ac.in}, Prasanjit Aich$^b$\footnote{prasanjita@iisc.ac.in}, Arkajyoti Bandyopadhyay$^a$\footnote{s22ph09003@iitbbs.ac.in}, and  \\ Sachindeo Vaidya$^b$\footnote{vaidya@iisc.ac.in}\\
\normalsize{\it $^a$ School of Basic Sciences, Indian Institute of Technology Bhubaneswar, Odisha 752050, India}\\
\normalsize{\it $^b$Centre for High Energy Physics, Indian Institute of Science, Bangalore, 560012, India}}

\begin{document}


%
\date{\today}

\maketitle

\begin{abstract}
Using variational methods, we numerically investigate the matrix model for the two-color QCD coupled to a single quark (matrix-QCD$_{2,1}$) in the limit of ultra-strong Yang-Mills coupling ($g =\infty$). The spectrum of the model has superselection sectors labelled by baryon number $B$ and spin $J$. We study sectors with $B=0,1,2$ and $J=0,1$, which may be organised as mesons, (anti-)diquarks and (anti-)tetraquarks.  For each of these sectors, we study the properties of the respective ground states in both chiral and heavy quark limits, and uncover a rich quantum phase transition (QPT) structure.  We also investigate the division of the total spin between the glue and the quark and show that glue contribution is significant for several of these sectors.  For the $(B,J)=(0,0)$ sector, we find that the dominant glue contribution to the ground state comes from reducible connections. Finally, in the presence of non-trivial baryon chemical potential $\mu$, we construct the phase diagram of the model. For sufficiently large $\mu$, we find that the ground state of the theory may have non-zero spin.

\end{abstract}


\section{Introduction} 

 $SU(2)$ gauge theory coupled to fundamental or adjoint quarks  (two-color QCD) has been the subject of considerable interest and investigation  for several decades now \cite{Nakamura:1984, Leutwyler:1992yt,Hands:1999md, Kogut:2000ek}.  Despite being quite different from the real world QCD, this theory, especially with $N_f$ quarks transforming in the pseudo-real representation of the gauge group (we will henceforth refer to this theory as QCD$_{2,N_f}$) has several novel and intriguing features which are worth studying in their own right. The pseudo-reality of the $SU(2)$ representation leads to an enlarged  $SU(2N_f)$ flavor symmetry -- the Pauli-G\"ursey symmetry \cite{Pauli:1957, Gursey:1958}, and  gives rise to to unusual  bound states like physical diquark states \cite{Splittorff:2000mm, Vanderheyden:2001gx, Wirstam:2002be, Splittorff:2002xn, Nishida:2003uj, Klein:2004hv, Fukushima:2007bj, Andersen:2010vu, Kanazawa:2020ktn }. Two-color QCD is of interest for lattice studies as well \cite{Hands:1999md,Kogut:2001na, Kogut:2001if, Kogut:2002cm,Muroya:2002ry,Kogut:2003ju,Braguta:2015cta,Astrakhantsev:2020tdl, Begun:2022bxj, Braguta:2023yhd, Iida:2024irv} since the determinant of the Euclidean Dirac operator is real, allowing the investigation of situations with finite baryon density \cite{Kogut:2000ek}.


In this article, we will study the matrix model version of the  QCD$_{2,1}$, which we will refer to as matrix-QCD$_{2,1}$. The matrix model, first proposed  for pure gauge theory in \cite{Balachandran:2014iya,Balachandran:2014voa} and with quarks in \cite{Pandey:2016hat, Pandey:2019dbp}, is simple to derive, but nontheless sophisticated enough to retain several non-trivial topological features of the full gauge field theory. For instance, it retains the information that the gauge bundle is twisted \cite{Singer:1978dk, Narasimhan:1979kf}, and when coupled to massless quarks, exhibits the chiral anomaly \cite{Acharyya:2021egi}. Since the model is quantum mechanical (rather than quantum field theoretic), it holds the promise to deliver non-trivial results in a rather straightforward manner. Indeed, it has led to fairly accurate predictions of the masses of glueballs as well as light hadrons \cite{Acharyya:2016fcn, Pandey:2019dbp}, which is rather surprising, given that the model is a drastic truncation of the full field theory.

Our primary tool of study is numerical and in this article, we will focus on the strong coupling limit of the theory.  Using variational methods, we will uncover properties of the 
low energy eigenstates, their energies, and the expectation values of some interesting and important observables. We will refer to these energy eigenstates as \textit{hadrons} of matrix-QCD$_{2,1}$. 

The Hamiltonian of matrix-QCD$_{2,1}$ commutes with the total (quark plus glue) \textit{spin}, corresponding to the spatial rotational symmetry $SO(3)_{rot}$. The usual $U(1)_V$ baryon number symmetry is enhanced to $SU(2)_B$ Pauli-G\"ursey symmetry.   Hence the hadrons can be arranged  in representations of $SO(3)_{rot}$ (labelled by  $J$) and $SU(2)_B$ (labelled by  $B$).  The physical Hilbert space can be decomposed into distinct $(B, J)$-sectors, which are superselected.  In our numerical simulations, we study the  low-energy regime of  the sectors with $(B,J)=(0,0), (0,1), (1,0), (1,1)$ and $(2,0)$.

Our  Hamiltonian (with a Dirac quark $Q$) naturally includes the term  $(c \bar{Q} \gamma^0 \gamma^5 Q)$  
originating in the curvature coupling of fermions on $S^3 \times \mathbb{R}$   \cite{Sen:1985dc}. Because  the chiral charge $\widehat{Q}_0\equiv \frac{1}{2}(\bar{Q} \gamma^0 \gamma^5 Q)$ is anomalous in the quantum theory \cite{Fujikawa:1979ay, Acharyya:2021egi}, $c$ should not be thought of as a chemical potential in the strict thermodynamic sense. Nonetheless, we have studied the effect of $c$ on the physics of the problem and find that there are first order phase transitions   in the chiral limit.  We emphasize that all these phase transitions are at zero temperature and hence are \textit{quantum phase transitions} (QPTs), arising from level crossing in the ground state. Interestingly, in different $(B,J)$ sectors, the QPT occurs at different critical values. We will denote the critical values in the $(B,J)=(0,0)$, $(1,1)$ and $(0,1)$ sectors as $c^\ast_0$, $c^\ast_1$ and $c^\ast_2$, respectively.


We can also compute the expectation values of quark spin and glue spin for the different $(B,J)$ hadrons. This provides a direct estimate of the contribution of the glue in the total spin of a state. Remarkably, we find that for several hadrons, the spin of the glue dominates the total angular momentum of the states, particularly in the chiral limit. On the other hand, the quark (almost always) contributes significantly to the total spin in the heavy fermion limit. 

We also study the model in the presence of the baryon chemical potential $\mu$. Although the baryon number $B_3$ commutes the rotations, we find that above a non-zero value $\mu^*$, the ground state of the model is not spin-0: rather it has spin-1. This behaviour is reminiscent of the LOFF phase \cite{Larkin:1964wok,Fulde:1964zz}, although in our case the ground state breaks rotational symmetry rather than the translational one.

In most theoretical discussions of quantum Yang-Mills theory, the gauge field is usually taken to be irreducible \cite{Atiyah:1978mv}. Irreducible connections have the technical advantage that their quotient by the group of gauge transformations has no non-trivial fixed points. However, it leaves open the question of the role and relative importance of reducible connections. Our investigations provide a tantalizing hint that for the $(B,J)=(0,0)$ sector, the dominant contribution at strong coupling comes from a class of reducible connections.  As we will show in some detail, the signature of this effect is imprinted in the the third and fourth Binder cumulants ($g_3$ and $g_4$).

This article is organized as follows.  In Section 2, we present the Hamiltonian for matrix-QCD$_{2,1}$ and its symmetries. Section 3 describes the numerical strategy for investigating the model in the strong coupling regime.  Section 4 and its subsections contain detailed description of the various physical features for each of the sectors with $(B,J)=(0,0)$, $(0,1)$, $(1,0)$, $(1,1)$ and $(2,0)$.   Specifically, we study the first order QPT in these sectors by focusing on the ground state energy, the chiral charge $Q_0$ and the fourth Binder cumulant $g_4$. We also investigate the division of the total spin between the quark and the glue.  Finally, we show that the {\it global} ground state  in the chiral limit is localized at (a class of) reducible connections. Our summary and outlook for future work are in Section 5.

\section{Matrix model of two-color QCD} 
The construction of the matrix model described in \cite{Balachandran:2014iya, Balachandran:2014voa}  involves the pullback of the Maurer-Cartan form of $SU(N)$ to an $S^3$ of radius $R$. The gauge field is described by a real 
rectangular matrix $M_{ia}$ with $i=1,2,3$ and $a=1,2\ldots (N^2-1)$, or equivalently the entries of 3 
hermitian matrices $\mathcal{A}_{i} \equiv M_{ia}T^a$, where $T^a$'s are the generators of $SU(N)$ in the fundamental representation. Under spatial rotations $R \in SO(3)_{rot}$, the gauge field 
transforms as  $\mathcal{A}_i  \to \mathcal{A}'_i = R_{ij} \mathcal{A}_j$, and and as 
$\mathcal{A}_i \to h^{-1} \mathcal{A}_i h$ under gauge transformations $h \in SU(N)$. The matrix degrees 
of freedom $M_{ia}$ are elements of  $\mathcal{M}(3,N)$ -- the set of all $3 \times (N^2-1)$-dimensional 
real matrices, and the configuration space of the pure Yang-Mills theory is the base space of the principle 
bundle $Ad \, SU(N) \to  \mathcal{M}(3,N) \to \mathcal{M}(3,N)/Ad \, SU(N) $. The chromomagnetic 
field $\mathcal{B}_{i}$ (or equivalently the gauge field curvature $\mathcal{F}_{ij} = \epsilon_{ijk}\mathcal{B}_{k}$) is given by $\mathcal{B}_{i} =- \frac{1}{R}\mathcal{A}_i-i \epsilon_{ijk}[\mathcal{A}_j,\mathcal{A}_k]$. It is convenient to work with dimensionless variables $M_{ia}' \equiv R M_{ia}$. For notational convenience we will henceforth drop the prime on the $M_{ia}$.

The quantum dynamics is based on $M_{ia}$ and their conjugate momenta $\Pi_{ia} = -i \frac{\partial \,\,\, }{\partial M_{ia}} $, the $\Pi_{ia}$ being the chromoelectric field.
Defining $\Pi_i \equiv \Pi_{ia} T^a$,  the standard Hamiltonian for Yang-Mills theory is 
\begin{eqnarray}
H_{YM} = \frac{1}{R} \left[\text{Tr}(\Pi_i \Pi_i + \mathcal{B}_{i}\mathcal{B}_{i})\right] = \frac{1}{R} \left[\text{Tr}\left(\Pi_i \Pi_i + \mathcal{A}_i \mathcal{A}_i + i g \epsilon_{ijk} [\mathcal{A}_i, \mathcal{A}_j] \mathcal{A}_k- \frac{g^2}{2} [\mathcal{A}_i, \mathcal{A}_j]^2 \right)\right],
 \end{eqnarray}
which we recognize as a multidimensional harmonic oscillator perturbed by cubic and quartic interactions. The glue Hilbert space is infinite-dimensional and the states are normalizable functions $f(M_{ia})$. The inner product on the states is defined with measure  $(dM_{11} dM_{12}\ldots dM_{33}) $, which is obviously invariant under spatial rotations and gauge transformations. 

It is both easy and straightforward to introduce quarks in the matrix model. Fundamental quarks are spinors $Q_{\alpha A} $ -- Grassmann-valued matrices which depends only on time \cite{Pandey:2016hat}, transforming in the spin-$\frac{1}{2}$ representation of spatial $SO(3)$ and in the fundamental representation of color $SU(2)$.  Here, $\alpha=1,2$ denotes the spin index and $A=1,2$ is the color index.  

The Dirac fermion $Q$ has a left Weyl fermion $b$ and a right Weyl fermion $c^\dagger$:
\begin{equation}
Q_{\alpha A} = \left(\begin{array}{c}
			b_{\alpha A} \\
			-i (\sigma_2)_{\alpha \beta} c^\dagger_{ \beta A} \end{array} \right) \equiv 
			\left(\begin{array}{c}
			b_{ \alpha A} \\
			d^\dagger_{\alpha A} \end{array} \right).
\end{equation}

The $b$'s and $d$'s satisfy the standard fermionic anti-commutation relations
\begin{equation}
\{ b_{\alpha A}, b^\dagger_{\beta B} \} = \delta_{\alpha \beta} \delta_{AB} = \{ d_{\alpha A}, d^\dagger_{\beta B} \} . 
\end{equation}
The fermionic Hilbert space of matrix-QCD$_{2,1}$ is finite dimensional: the maximum number of fermions that can be accommodated in the states is eight. $|0_F\rangle$ denotes the $0-$fermion state satisfying $ b_{\alpha A } |0_F\rangle =0 = d_{\alpha A} |0_F\rangle$ and $|8_F\rangle$ is the $8-$fermion state with $ b^\dagger_{\alpha A } |8_F\rangle =0 = d^\dagger_{\alpha A} |8_F\rangle$ for all $(\alpha,A)$. 

In the Weyl basis, the Hamiltonian for minimally coupled  massive quark is given by  \cite{Pandey:2016hat}
\begin{eqnarray}
 H &=&  H_{YM} + \frac{1}{R}\left( \tilde{c} H_c + m \, H_m + gH_{int}\right), \label{Hamiltonian_1}
 \end{eqnarray}
 where the curvature coupling, quark mass and the quark-gluon interaction terms are given by 
  \begin{eqnarray}
&& H_{int} =  M_{ia} (b^\dagger_{\alpha A } \sigma^{i}_{\alpha \beta}T^a_{AB}  b_{\beta B } - d_{\alpha A f} \sigma^{i}_{\alpha \beta} T^a_{AB}  d^\dagger_{\beta B}), \nonumber \\ 
&& H_c = b^\dagger_{\alpha A} b_{\alpha A} - d_{\alpha A}  d^\dagger_{\alpha A }, \quad {\rm and}  \quad \\
&&m H_m = m\left( e^{i \theta}\, b^\dagger_{\alpha A}d^\dagger_{\alpha A} +e^{-i \theta}\, d_{\alpha A}  b_{\alpha A }\right) = (m \cos \theta)\, H_m^R+ i (m \sin \theta) H_m^I. \nonumber 
\end{eqnarray}

The dimensionless Yang-Mills coupling constant is $g$, while $R^{-1}$ determines the energy scale of the system. Here, we consider a complex (dimensionless) mass term for the quark, whose magnitude is  $m\in \mathbb{R}_+$ and $e^{i\theta} $  its phase. The presence of the imaginary mass term $(\theta \neq n \pi)$ can be justified as follows: in general, the Yang-Mills-Dirac Lagrangian may have the term  $\sim \theta  (F\wedge F)$. This term can be eliminated by an chiral $U(1)_A$ rotation, but at the cost of generating the imaginary mass term.  In the matrix model, the $U(1)_A$ rotations are generated by the chiral charge
\begin{eqnarray}
  \widehat{Q}_0 \equiv \frac{1}{2} \left(b^\dagger_{\alpha A} b_{\alpha A} - d_{\alpha A}  d^\dagger_{\alpha A }\right),\label{charge_2}
\end{eqnarray}
which indeed commutes with the Hamiltonian if the quarks are massless.  In the quantum theory, $U(1)_A$  is explicitly broken when $m \neq 0$ and is anomalously broken when $m=0$.  As a result, theories with different  $\theta$ correspond to different physical systems. 


It is straightforward to verify that $H_m^R$, $H_m^I$ and $2\widehat{Q}_0$ generate an $SU(2)$, and hence it is natural to treat the parameters $m \cos \theta$, $m \sin \theta$ and $ \widetilde{c}$ on a democratic footing.

Under spatial rotations, the glue and the quark transform in spin-1 and spin-$\frac{1}{2}$ representations respectively. Their generators are
\begin{eqnarray}
L_i\equiv -2\epsilon_{ijk}  \text{Tr}\left(\Pi_j \mathcal{A}_k \right), \quad \quad S_i\equiv \frac{1}{2} (b^\dagger_{\alpha A} \sigma^i_{\alpha \beta } b_{\beta A} +d_{\alpha A} \sigma^i_{\alpha \beta } d^\dagger_{\beta A}), 
\label{spin_operator_1}
\end{eqnarray}
and satisfy $[L_i, L_j]=i\epsilon_{ijk} L_k$, and $  [S_i, S_j]=i\epsilon_{ijk} S_k. $
The Hamiltonian does not commute with $L_i$ and $S_i$ separately, but commutes with $J_i =L_i +S_i$. The $J_i$'s obey  $[J_i, J_j]=i\epsilon_{ijk} J_k$. 

Under a color $SU(2)$ rotation, the glue transforms as a vector, while the quark transforms in the fundamental representation. The gauge rotations of glue and quarks in the color space are generated by 
\begin{equation}
G^a_{glue} \equiv 2i \text{Tr}\Big([\Pi_i, \mathcal{A}_i] T^a\Big), \quad \quad  G^a_{quark} \equiv   (b^\dagger_{\alpha A} T^a_{AB} b_{\alpha B} +d_{\alpha A} T^a_{AB} d^\dagger_{\alpha B}).
\end{equation}
The Hamiltonian commutes with $G^a = G^a_{glue} + G^a_{quark}$ which satisfy $[G^a, G^b]=i \epsilon^{abc} G^c$. The operators $G^a$'s are the generators of the Gauss' law constraint.

\subsection{Strong coupling regime} 
In this article, we are interested in the  large $g$ (strong coupling) regime.  To this end, it is convenient to rescale $\mathcal{A}_i \to g^{-\frac{1}{3}}\mathcal{A}_i$ and $\Pi_i \to g^{\frac{1}{3}}\Pi_i$. Further, let us define $c \equiv \tilde{c} g^{-\frac{2}{3}} $, $M \equiv m g^{-\frac{2}{3}} $  and $e_0\equiv g^{\frac{2}{3}} R^{-1}$. The Hamiltonian (\ref{Hamiltonian_1}) then takes the form  
\begin{eqnarray}
H \equiv e_0 \left[ \text{Tr}\left(\Pi_i \Pi_i -\frac{1}{2} [\mathcal{A}_i, \mathcal{A}_j]^2 + g^{-\frac{4}{3}} \mathcal{A}_i \mathcal{A}_i + i g^{-\frac{2}{3}} \epsilon_{ijk} [\mathcal{A}_i, \mathcal{A}_j] \mathcal{A}_k \right)+ c H_c+ M H_m+ H_{int}\right], \label{strong_coupling_Hamiltonian}
\end{eqnarray}
where $e_0$ determines the energy scale of the system. In the double scaling limit $g, R \to \infty$, the Hamiltonian has a meaningful spectrum provided $e_0$ is held finite.

In this double scaling limit, we see from (\ref{strong_coupling_Hamiltonian}) that the quadratic and cubic gluon self-interaction terms are sub-dominant. Thus in the large $g$ limit with fixed $c$ and $M$, all eigenvalues and eigenvectors of $(e_0^{-1}H)$ are independent of $g$.

The term  $ c H_c$ may be thought of  as the ``chiral chemical potential''. Of course, chiral symmetry is broken anomalously and hence, this term is not a chemical potential in the thermodynamic sense \cite{Braguta:2015owi,Braguta:2016aov }. Nonetheless, it is interesting to study the effect $c$ on the energy spectrum and expectation values of other interesting observables. We explore the situations when $c \in [0,1.5]$, which is the region where all interesting physics happens.

When $M \simeq 0 $, the effect of the fermion mass is negligible: this is the chiral limit. On the other hand, $M \gg 0$ corresponds 
to the heavy quark limit. Here, in our numerical diagonalization of $H$ in (\ref{strong_coupling_Hamiltonian}), we find it is sufficient to study $M \in [0,4]$ to investigate both these limits. 

\subsection{Symmetries of the Hamiltonian and the physical Hilbert space  } 
In the matrix-QCD$_{2,1}$, the $U(1)_V$  baryon number symmetry is enhanced to $SU(2)_B$, whose generators 
\begin{eqnarray}
\widehat{B}_1 &=& \frac{1}{2} \epsilon_{\alpha \beta} \epsilon_{AB} ( b^\dagger_{\alpha A} d_{\beta B} +d^\dagger_{\alpha A} b_{\beta B}),  \nonumber  \\ 
 \widehat{B}_2 &=&  \frac{i}{2} \epsilon_{\alpha \beta} \epsilon_{AB} ( d^\dagger_{\alpha A} b_{\beta B}-b^\dagger_{\alpha A} d_{\beta B} ),  \label{charge_1}  \\ 
\widehat{B}_3 &=&  \frac{1}{2} (b_{\alpha A } ^\dagger b_{\alpha A } -d_{\alpha A} ^\dagger  d_{\alpha A })\nonumber 
\end{eqnarray}
satisfy 
\begin{eqnarray}
&& [\widehat{B}_p, \widehat{B}_r] =i \epsilon_{prs} \widehat{B}_s, \quad  
[\widehat{B}_p, \widehat{Q}_0] =0, \quad [H, \widehat{B}_p] =0. 
\end{eqnarray}


 When the quarks are massive, the $U(1)_A$ symmetry is explicitly broken to $\mathbb{Z}_2$.   But when the quarks are massless the Hamiltonian only formally commutes with $\widehat{Q}_0$, and  the quantum theory only preserves a $\mathbb{Z}_2 \subset U(1)_A$ symmetry due to chiral anomaly \cite{Fujikawa:1979ay, Acharyya:2021egi}.

On inclusion of the baryon chemical potential term in the Hamiltonian $H_\mu = e_0 \mu \widehat{B}_3$, the $SU(2)_B$ symmetry is explicitly broken since $[H_\mu, \widehat{B}_1]\neq 0 \neq [H_\mu, \widehat{B}_2]$. Thus for $\mu \neq 0$, the global symmetry of matrix-QCD$_{2,1}$ is reduced to $U(1)_B \times \mathbb{Z}_2$. Note that, $H_\mu$ does not break the spatial rotational symmetry.  In section \ref{chem_pot_sec}, we will discuss the consequences  of the baryon chemical potential on  the hadron spectra of matrix-QCD$_{2,1}$. 

The total Hilbert space $\mathcal{H}$ is the direct product space $\mathcal{H}_F \otimes \mathcal{H}_G$ where $\mathcal{H}_F$ and $\mathcal{H}_G$ contain the quark and glue states respectively. 
The Gauss law constraint demands that all physical observables  commute with $G^a$. 
The physical Hilbert space $\mathcal{H}_{phys}$ contains vectors which are annihilated by the Gauss generators: $G^a | \Psi \rangle =0$ for all $| \Psi \rangle \in \mathcal{H}_{phys}$ \cite{Pandey:2019dbp}. 

 $\mathcal{H}_{phys} $ can be spanned by colorless eigenstates of the Hamiltonian. Because $H$ commutes with $J_i$  and $\widehat{B}_p$, the color-singlet energy eigenstates can also chosen to be  simultaneous eigenstates of $(J_iJ_i)$,  $J_3$, $(\widehat{B}_p \widehat{B}_p)$ and $\widehat{B}_3$.  Such a color-singlet energy eigenstate  
 is  labelled by the quantum numbers $\{J,J_3, B, B_3\}$ where $J(J+1)$ and $B(B+1)$ are the eigenvalues of  $(J_iJ_i)$ and $(\widehat{B}_p \widehat{B}_p)$, respectively. 
The colorless energy eigenstates necessarily have an even number of quarks: a state with an odd number of quarks carry half-integer color-charge, which cannot combined with any glue state to produce a color-singlet state.  
The fermionic Hilbert space is finite dimensional and  it is easy to see that the physical states can have max$(B)=2$.  

The glue states and the states with even number of quarks  transform in the integer representation of the algebra generated by $L_i$ and $S_i$, respectively.  Consequently, the colorless eigenstates of the Hamiltonian always transform in an integer representation of  $SO(3)_{rot}$: $J=0,1,2 \ldots$ 
Further, states with even number of fermions can be of three distinct types: 
\begin{enumerate}[i)]
\item mesons: these have equal number of quarks and anti-quarks and have $B_3=0$. 
\item  diquarks/anti-diquarks: these are states where the number of quarks and anti-quarks differ by two and have $B_3= \pm1$.  
\item tetraquarks/anti-tetraquarks: these are states where the number of quarks  and anti-quarks  differ by four and have $B_3= \pm 2$.
\end{enumerate}
Thus for the colorless energy eigenstates, $J=0,1,2\ldots$ and  $B=0,1,2$.

\section{Numerical diagonalization of the Hamiltonian}

We are interested in the energy eigenstates in the double scaling limit: the low-energy dynamics in this deep non-perturbative regime leads to bound states of quarks and gluons, which are identified with the low-lying hadrons. In this section, we construct the  variational energy eigenstates of  matrix-QCD$_{2,1}$ in the  $g^{-2/3} (\equiv \nu)=0$ limit. 

The fermionic Hilbert space $\mathcal{H}_F$ is finite-dimensional and the space of states with even number of fermions contains 128 linearly independent fermionic states $\{|\psi_{n_F}\rangle \, :\, n_F=1,2,\ldots 128\}$. These fermionic states have spin-0, spin-1 or spin-2, and transforms in singlet, triplet or quintuplet representations of color $SU(2)$. We can expand the fermionic part of the color-singlet energy eigenstates in the basis of $\{|\psi_{n_F}\rangle \}$.

The bosonic Hilbert space  $\mathcal{H}_G$ is infinite-dimensional and we need  a basis for $\mathcal{H}_G$ to expand the bosonic part of the color-singlet energy eigenstates.   We can use the set of eigenstates of $H_{ref}$ -- the Hamiltonian of a nine-dimensional harmonic oscillator 
\begin{eqnarray}
H_{ref} = \frac{1}{2} \Pi_{ia} \Pi_{ia} + \frac{1}{2}  M_{ia} M_{ia}
\end{eqnarray}
as the basis for  $\mathcal{H}_G$. Let $\{|\phi_{n_b}^d\rangle: \, n_b=0,1,2 \ldots \}$ be the eigenstates of $H_{ref}$: 
\begin{eqnarray}
H_{ref} |\phi_{n_b}^d\rangle = \Big( n_b +\frac{9}{2} \Big)  |\phi_{n_b}^d\rangle.
\end{eqnarray}
The $d$, which depends on $n_b$, is a multi-index object accounting for the degeneracy for a given $n_b$. 

The composite states $|\psi_{n_F}\rangle \otimes |\phi_{n_b}^d\rangle$ can transform in various representations of the gauge group and can have different $J$ and $B$. Here, we are interested in constructing a basis for  $\mathcal{H}_{phys}$ -- the space of colorless states. So we only need those composite states $\{|\Phi_{n_F, n_b}^d \rangle \} \subset \{ |\psi_{n_F}\rangle \otimes |\phi_{n_b}^d\rangle\}$ which transform in the singlet representation of color $SU(2)$.

Any normalized energy eigenstate $| \Psi \rangle $ can be expanded as 
\begin{eqnarray}
| \Psi \rangle = \sum_{n_b=1}^\infty \sum_d \sum_{n_F=1}^{128} C_{n_F, n_b}^d |\Phi_{n_F, n_b}^d \rangle, \quad\quad \sum_{n_b=1}^\infty \sum_d \sum_{n_F=1}^{128} |C_{n_F, n_b}^d|^2=1.  \label{ansatz}
\end{eqnarray}
Solving the Schr\"odinger equation means finding the coefficients $C_{n_F, n_b}^d$ and the energies $E$. 
Since the $H $ in (\ref{strong_coupling_Hamiltonian})  is a multi-dimensional quartic oscillator, $C_{n_F, n_b}^d $'s cannot be determined  exactly by analytic methods. Here we adopt the Rayleigh-Ritz method to estimate the $C_{n_F, n_b}^d$'s and the $E$'s.   To this end, we will truncate the summation in \ref{ansatz} with $n_b \leq N_{max}$. 


In any numerical work, a central question is convergence of the data. We will progressively increase $N_{max}$ till we reach convergence in energy eigenvalues (and other observables). Specifically, we have found that $N_{max}=16$ is sufficient to achieve excellent convergence.  
 In all the computations presented here, we have taken $N_{max} = 18$.

%
%
%
%
%
%

The convergence of the variational calculation is substantially improved if we use the (isotropic) squeezed oscillator 
\begin{eqnarray}
H_{x} = \frac{x}{2} \Pi_{ia} \Pi_{ia} + \frac{1}{2x}  M_{ia} M_{ia}
\end{eqnarray}
instead of $H_{ref}$ and further extremize the ground state energy with respect to the  squeeze parameter $x$.  This is computationally more expensive but the reward in terms of the improved convergence is significant. 

Details of convergence of the energy eigenvalues is discussed in appendix \ref{app:convergence}.

\section{Results} \label{results_sec}

\begin{figure}[t!]
\begin{center}
\includegraphics[width=16cm]{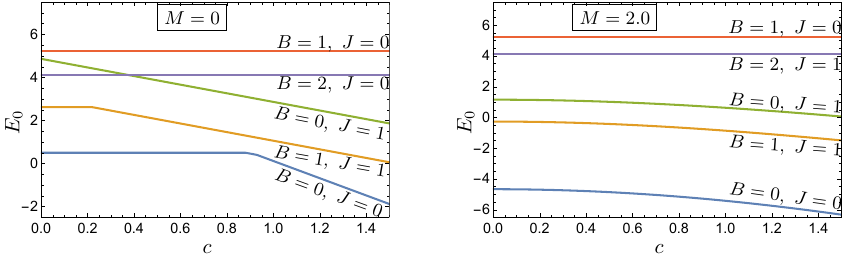}
\vspace*{-0.5cm}
\caption{Energies of the ground states in the low-lying sectors.}\label{Fig_1}
\end{center}
\end{figure}

Using the strategy discussed above, we obtain the low-lying spectrum of spin-0 and spin-1 hadrons (see Fig.~\ref{Fig_1}). The space of these low-lying hadrons  splits into five disjoint subspaces characterized by their values of  baryon number $B$ and spin $J$. The spin-0 hadrons can have $B=0,1,2$, while spin-1 hadrons have $B=0$ or 1. 
Irrespective of the quark mass, the lightest hadron (global ground state) has the quantum numbers $(B,J) =(0,0)$, while the lightest spin-1 hadron has $(B,J) =(1,1)$. 

In the subsections that follow, we present the expectation values of the scalar observables like Hamiltonian $H$ (specifically the ground state energy), chiral charge $Q_0$, quark and glue spin $L^2$ and $S^2$ in every sector.  In sectors with non-zero $J$, the expectation values of the vector quantities like $L_3$ and $S_3$ help us to understand the distribution of the total spin between the quark and the glue.

 Another pair of useful quantities for our study are
\begin{eqnarray}
 g_3 \equiv \frac {\sqrt{3} }{2} \frac{\epsilon_{ijk} \epsilon_{abc} \langle M_{ia} M_{jb} M_{kc}\rangle  } {\langle M_{ia} M_{ia}\rangle ^{\frac{3}{2}}}, \quad\quad 
g_4 \equiv \frac{9}{8} \left[\frac{ \langle  M_{ia} M_{ib} M_{ja} M_{jb}  \rangle }{\langle  M_{ia} M_{ia}\rangle ^{2}} - \frac{1}{2} \right].
\end{eqnarray}
These are like the third and fourth Binder cumulants which are extensively used in study of spin systems \cite{Binder:1981}.



We find that in the chiral limit, varying $c$ reveals first order QPTs in the $(B,J)=(0,0)$, $(1,1)$ and $(0,1)$ sectors. These QPTs occur because of level-crossings in ground state of these sectors. They are located at critical point $c^\ast_0\simeq 0.928$ for $(B,J)=(0,0)$, at  $c^\ast_1\simeq0.22$ for  $(B,J)=(1,1)$ and at $c_2^\ast=0$ for $(B,J)=(0,1)$. Noticeably,  $c^\ast_0$, $c^\ast_1$ and $c^\ast_2$ are all different.

 The first order nature of these QPTs is captured by the ground state expectation values of the chiral charge $Q_0$, which can be computed for each sector.  These QPTs are also validated by discontinuities in the fourth Binder cumulant at the critical points. 

We can also estimate the contribution of the quark and glue to the total spin of the hadron. In low energy QCD, this question in context of proton has been a subject considerable discussion since the first experimental measurement of the spin of proton estimated the quark contribution to be 
 $\sim4-24\%$ \cite{EuropeanMuon:1987isl}.
 Subsequent experiments 
estimate of the quark contribution  $\sim 33\%$ of the total spin \cite{COMPASS:2006mhr, HERMES:2006jyl}. Here in matrix-QCD$_{2,1}$, we present the estimates for the contribution of the quark to the spin of hadrons belonging different $(B,J)$ sectors. Generally we find that in the chiral limit, the quark spin is swallowed up by the glue. For spin-1 states, this effect is clearly observed in the expectation values of $L_3$ and $S_3$. For spin-0 states, this is indirectly visible in the expectation values of $L^2$ and $S^2$.

This division of spin between the quark and glue is interesting in both the chiral and heavy quark limits, which we demonstrate sector-wise. 

Finally, by examining the third and fourth Binder cumulants in one of the phases, we find  that for the global ground state in chiral limit, the gauge configurations are overwhelmingly reducible connections. 

\begin{figure}[h!]
\begin{center}
\includegraphics[width=16cm]{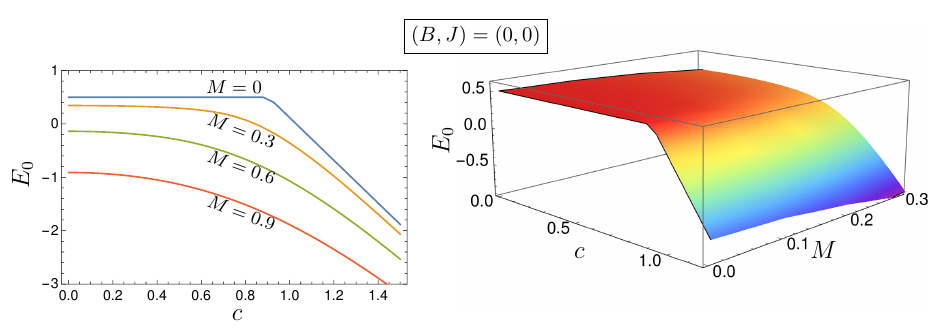}
\vspace*{-0.5cm}
\caption{Ground state energy as function of $c$ and $M$ for the $(B,J)=(0,0)$ sector.} \label{Fig_2}
\end{center}
\end{figure}

\subsection{$(B,J)=(0,0)$ sector }

Fig.~\ref{Fig_2} shows that when the quark mass $M$ vanishes, there is a level crossing in the ground state at $c=c^\ast_0$.  A closer inspection reveals that this level crossing  is rather very special because its is an accidental \textit{triple} crossing (i.e. the three lightest energy levels cross at $c=c^\ast_0$ when $M=0$). Typically, accidental degeneracies in interacting systems can only be uncovered numerically. 

\begin{figure}[h!]
\begin{center}
\includegraphics[width=16cm]{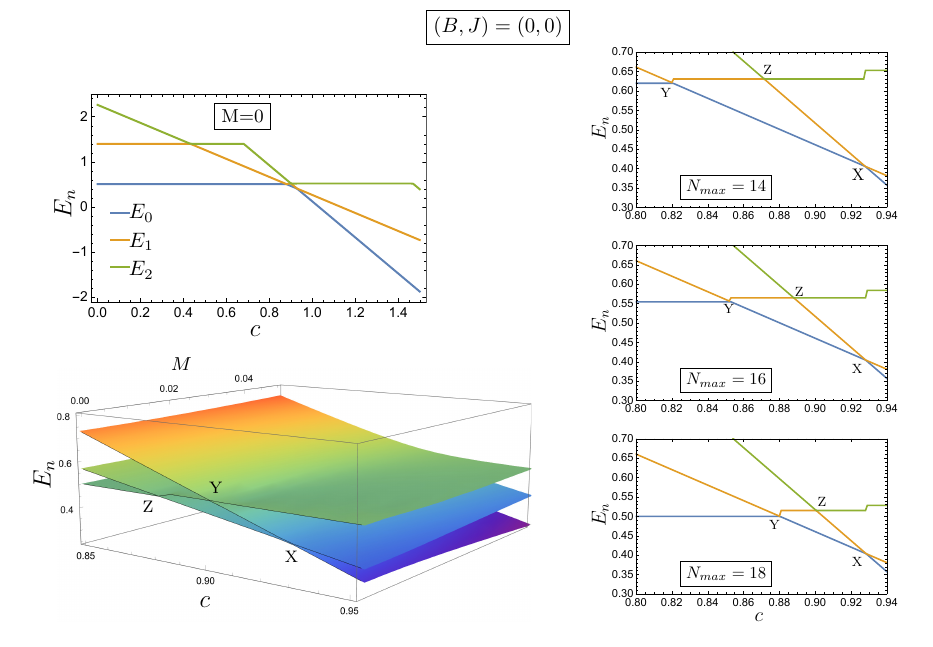}
\vspace*{-0.5cm}
\caption{ The triple crossing in the chiral ground state of the $(B,J)=(0,0)$ sector. Top left: The first three energy levels vs  $c$ at $M=0$. Bottom left:  The first three energy levels vs  $c$ at $M=0$. Right: The first three energy levels vs  $c$ at $M=0$ in the vicinity of the critical point for $N_{max}=14,16,$ and $18$. } \label{Fig_3}
\end{center}
\end{figure}
To  test the robustness of the numerical evidence for the triple crossing, we  plot the first three energy levels in the vicinity of the critical point for different values of $N_{max}$.  For finite $N_{max}$, the three levels do not intersect exactly at the same point:  rather they form a ``triangle'', as shown in Fig.~\ref{Fig_3}. As we progressively increase $N_{max}$, the point X of the triangle  remains fixed, while  Y and  Z move toward X as the triangle shrinks. Thus in the converged limit, the triangle is replaced just by point X, which we can confidently identify as the location of the triple crossing. Normally,  locating the exact value of the crossing usually requires a more sophisticated treatment like finite-size scaling. Here, we are fortunate that point X does not change with $N_{max}$ and we readily obtain its location: $c^\ast_0 \simeq 0.928$.

\begin{figure}[h!] \hspace*{-0.25 cm}
\begin{center}
\includegraphics[width=17cm]{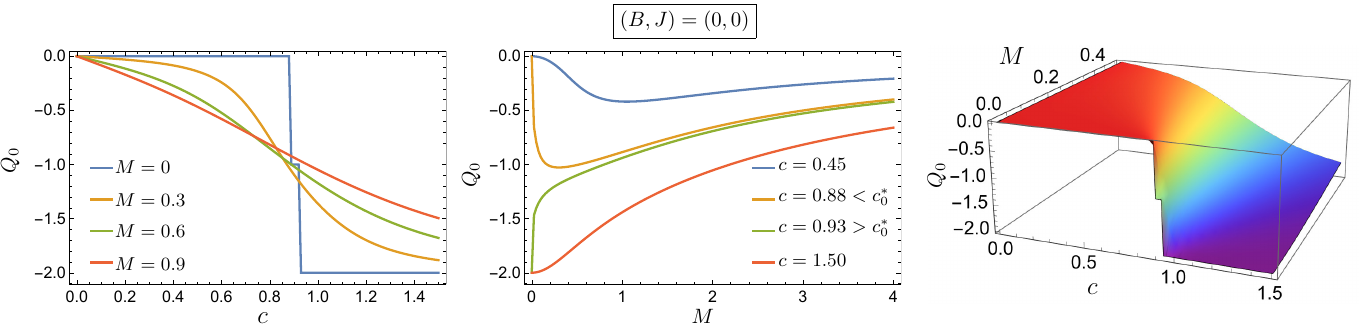}
\vspace*{-0.25cm}
\caption{ $Q_0$ in the ground state as a function of $c$ and $M$ for $(B,J)=(0,0)$.} \label{Fig_4}
\end{center}
\end{figure}

The triple crossing in the ground state in the chiral limit corresponds to a QPT at $c^\ast_0 \simeq 0.928$. The properties of this QPT are captured in the ground state expectation values of several observables. Among these, the most crucial is the chiral charge $  Q_0 $ (shown in Fig.~\ref{Fig_4}) which displays a sharp discontinuity at $c^\ast_0 \simeq 0.928$ when $M=0$ . Since $\langle Q_0 \rangle =( \partial E_0/\partial c)$,  the discontinuity in  $Q_0 $ captures the first order nature of the QPT. 

\begin{figure}[t!]
\begin{center}\hspace*{-0.4cm}
\includegraphics[width=16cm]{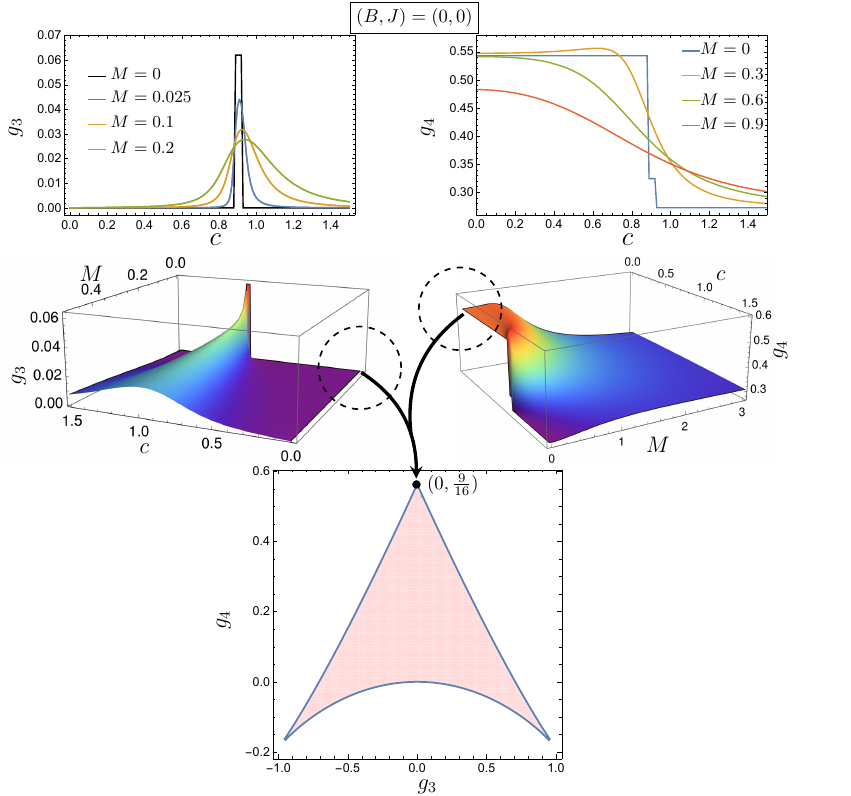} 
\caption{$g_4$ and $g_3$ as functions $c$ and $M$  for  $(B,J)=(0,0)$. The red shaded region represents the allowed values of $g_3$ and $g_4$ for all possible gauge field configurations (for details see \cite{Pandey:2016hat}). } \label{Fig_6}
\vspace*{-0.5cm}
\end{center}
\end{figure}

This first order QPT also induces a discontinuity at $c^\ast_0$ in the fourth Binder cumulant $g_4$ as shown in Fig.~\ref{Fig_6}.

 The third Binder cumulant $g_3$ is shown in Fig.~\ref{Fig_6}. We observe that $g_3$ looks like a  ``wall'' of finite height at the critical point $(c,M)=(c^\ast_0,0)$. With increasing $N_{max}$, the height of the wall remains the same while its width decreases. Slightly away from the critical point, $g_3 =0$ in both phase. Looking at the variation of $g_3$ with $M$ near the critical point (see Fig.~\ref{Fig_6}),  we can see that as $M$ decreases, the height of the bump near $c_0^\ast$ increases and it becomes an infinitesimally thin wall in the $M\to 0$ limit.

\begin{figure}[h!] 
\begin{center}
\includegraphics[width=17cm]{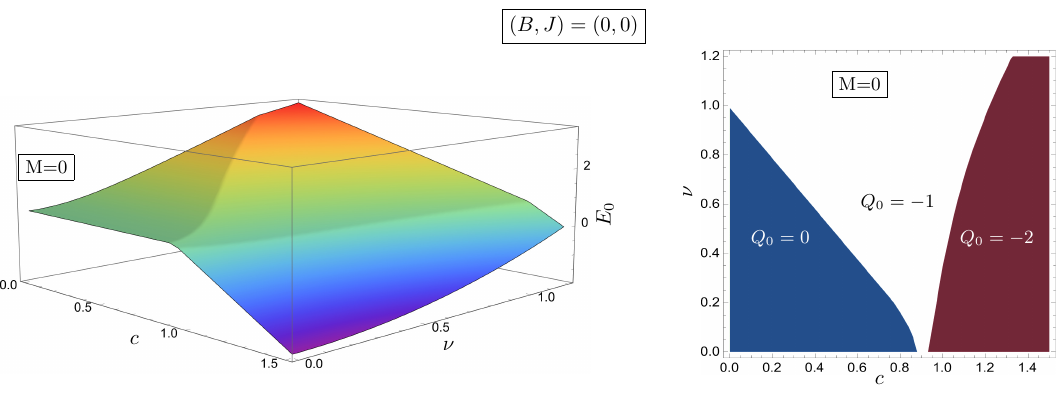} 
\vspace*{-0.5cm}
\caption{ $E_0$ and $Q_0$ in the ground state at $M=0$ for various $c$ and $\mu$. There are two first order QPTs for non-zero $\nu \equiv g^{-2/3}$.} \label{Fig_5}
\end{center}
\end{figure}

A deeper understanding of the triple crossing comes from examining the system at non-zero values of $\nu\equiv g^{-2/3}$.  As shown in Fig.~\ref{Fig_5}, for small non-zero $\nu$, the system shows two distinct first order QPTs (and three different phases) in the chiral limit.    As $\nu$ decreases, the two transition lines comes closer and  merge at the \textit{triple point} when $\nu=0$. 

When the quark mass is non-zero, the sharp discontinuity in  $\langle Q_0 \rangle $ and $g_4$ is softened and the phase transition at $M=0$ gives way to a continuous crossover.

\begin{figure}[t!]
\begin{center}
\includegraphics[width=17cm]{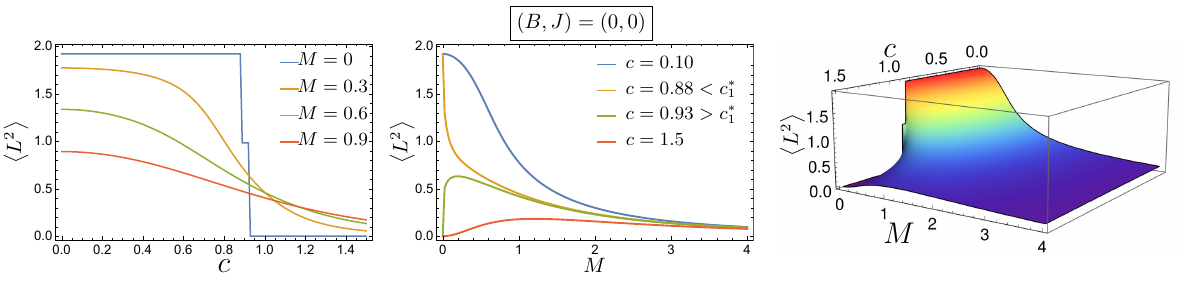}
\vspace*{-0.5cm}
\caption{$\langle L^2 \rangle $ as a function of $c$ and $M$ for $(B,J)=(0,0)$.} \label{Fig_7}
\end{center}
\end{figure}

While the total spin of the ground state in $(B,J)=(0,0)$ sector is zero, it can have contributions from  glue and quark states with higher angular momenta ($L$ and $S$). 
It is clear from  Fig.~\ref{Fig_7} that in the chiral limit 
\begin{eqnarray}
\lim_{M\to 0} \langle L^2 \rangle  &\simeq& \left\{\begin{array}{ll}
1.9 & \quad \text{for } c< c^\ast_0 \\
0 &\quad \text{for } c>c^\ast_0.
\end{array}\right.
\end{eqnarray} 
In other words, the chiral ground state in the phase with $c< c^\ast_0$ is dominated by spin-1 glue (and spin-1 quark) states, while for $c>c^\ast_0$, the ground state  is entirely composed of a spin-0 color-singlet glue (and quark) states. Since the values of $\langle S^2\rangle$ and $\langle L^2\rangle$ are identical in this sector,  we have not shown $\langle S^2\rangle$ separately.

For small $M$, the dominance of spin-1 glue states  for $c< c^\ast_0$ persists. However, as the value of $M$ increases, the contribution of the glue to the spin decreases for $c< c^\ast_0$. For  $c>c^\ast_0$ and small $M$, the contribution of the glue initially increases. But as the mass becomes significant, the glue contribution to the total spin starts to decrease. In the heavy quark limit $M\gg 0$, the contribution of the glue  to the total hadron spin becomes very small, irrespective of the value of $c$. 


In the phase with $M=0$ and  $c> c^\ast_0$, the ground state has $Q_0 =-2$ and hence is a state of the form $|0_F\rangle \otimes |\Psi_0(\mathcal{A}_i)\rangle$, where $|0_F\rangle$ is the 0-fermion state and $|\Psi_0(\mathcal{A}_i)\rangle$ is the spin-0 color-singlet state of glue. Thus in this phase, the ground state is separable and  is essentially a scalar glueball.

In contrast, in the ground state for $M=0$ and $c< c^\ast_0$, the quarks and glue are entangled. A closer inspection of $g_3$ and $g_4$ in this phase show that $g_3 \simeq 0$ and $g_4 \simeq 9/16$. In the $g_3-g_4$ plane,  the point $(g_3, g_4) = (0,9/16)$ is a very interesting because it corresponds to gauge fields $M_{ia}$ with only one non-zero singular  value \cite{Pandey:2016hat}.   In Fig.~\ref{Fig_6}, this point is the top tip of the red shaded region (the ``arrowhead'').  Thus, the ground state in this phase is dominated by glue states with reducible gauge field configurations.

\subsection{$(B,J)=(1,1)$ sector} 
\begin{figure}[h!]
\begin{center}
\includegraphics[width=16cm]{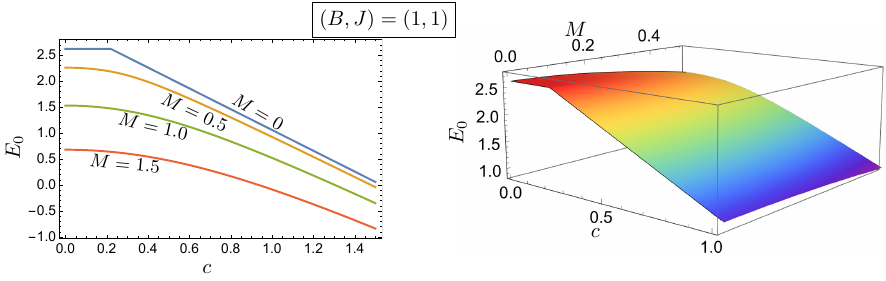}
\vspace*{-0.5cm}
\caption{ Energy $E_0$ of the lightest state for $(B,J)=(1,1)$ as a function of $c$ and $M$ .} \label{fig_8}
\end{center}
\end{figure}

Here the hadrons are spin triplets with $J_3=0,\pm 1$.  Scalar observables like $H, Q_0$, $L^2$, $S^2$, $g_3$ and $g_4$ do not distinguish between different values of $J_3$. Vector observables like $S_3$ and $L_3$ do depend on the $J_3$-value and we will denote their expectation values  for $J_3=0, \pm 1$ as $\langle \cdots \rangle_{J_3}$.  We investigate the behaviour of these observables as functions of $c$ and $M$.

\begin{figure}[h!]
\begin{center}
\includegraphics[width=18cm]{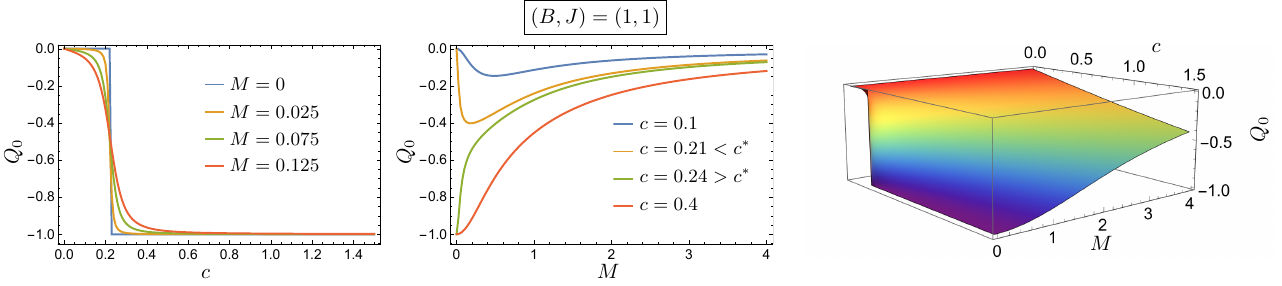}
\vspace*{-0.5cm}
\caption{Chiral charge $Q_0$  as a function of $c$ and $M$ for $(B,J)=(1,1)$.} \label{fig_9}
\end{center}
\end{figure}

\begin{figure}[h!]
\begin{center}
\includegraphics[width=16cm]{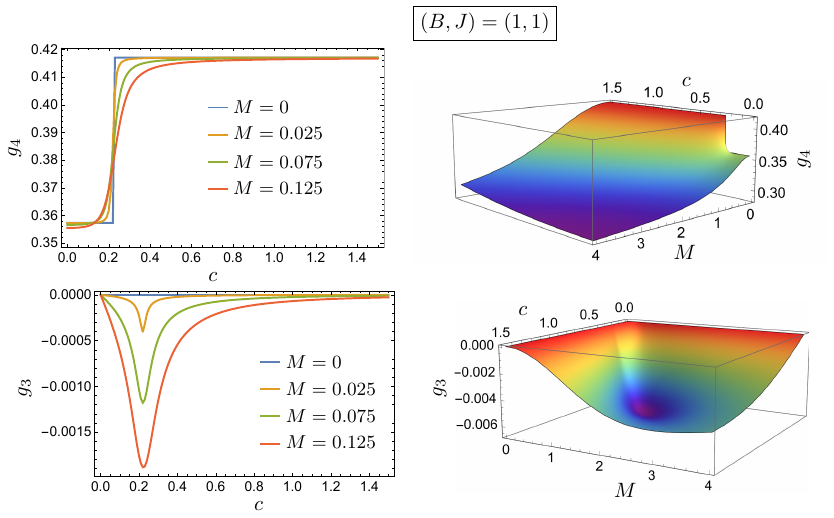}
\vspace*{-0.5cm}
\caption{$g_4$ and $g_3$  as functions of $c$ and $M$ in the $(B,J)=(1,1)$ sector} \label{fig_10}
\end{center}
\end{figure}

Here too, in the chiral limit, there is a level crossing as we vary $c$, leading to a first order QPT at $c^\ast_1 \approx 0.22$ (see Fig.~\ref{fig_8}).  The  phase transition at $(c,M)=(c^\ast_1,0)$ is further confirmed by plotting  $\langle Q_0 \rangle$ and $g_4$, both of which display a sharp discontinuity at $c_1^\ast$, as shown in Figs. \ref{fig_9} and \ref{fig_10}. 

For small non-zero $M$, the abrupt discontinuous jumps at $c^\ast_1$ get smoothened, and the first order transition is replaced by continuous crossover. Noticeably,  $g_3$ remains zero for all $c$ in the chiral limit and  shows an interesting dependence on small mass in the vicinity of $c^\ast_1$.

\begin{figure}[t!]
\begin{center}
\includegraphics[width=17cm]{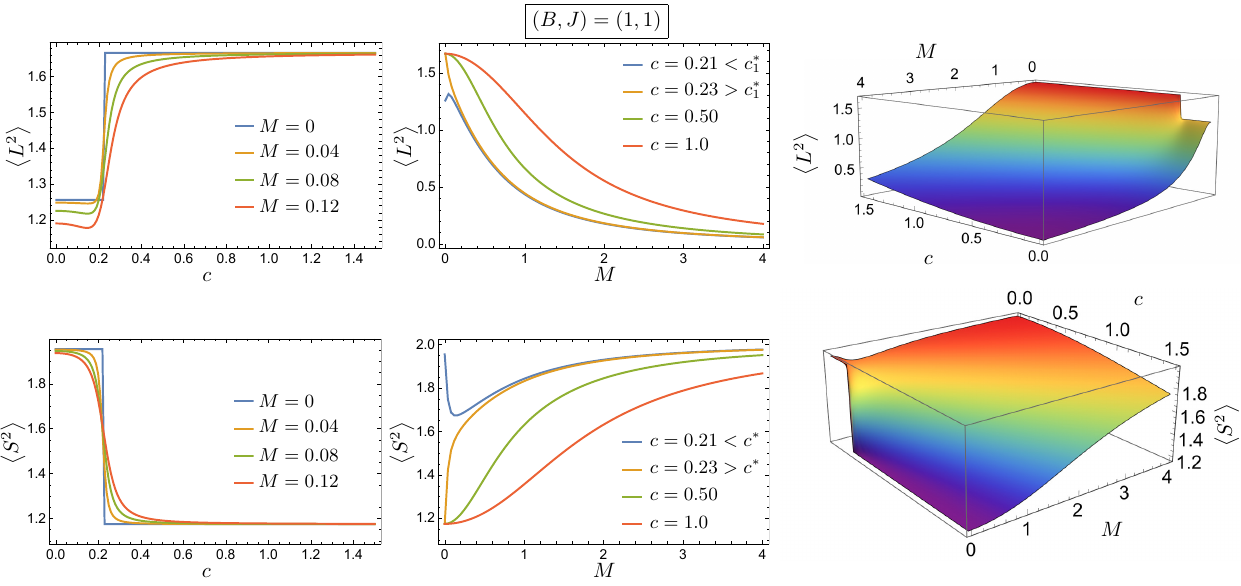}
\vspace*{-0.25cm}
\caption{$\langle L^2 \rangle $ (top) and $\langle S^2\rangle $ (bottom) as a function of $c$ and $M$  for $(B,J)=(1,1)$.} \label{fig_11}
\end{center}
\end{figure}

In this sector too, we can study the distribution of the total spin between the quark and the glue. Here, we find that the situation is even more dramatic.  For the lightest spin-1 triplet (i.e. $J^2 =1(1+1)$),  $\langle L^2 \rangle$ and $\langle S^2 \rangle $  are shown in Fig.~\ref{fig_11}.  Both $\langle L^2\rangle$ and $\langle S^2\rangle$ jump abruptly at $c_1^\ast$ in the chiral limit.  As it is evident from Fig.~\ref{fig_11}, for $c < c^\ast_1$, the dominant contribution to the total spin comes from quark, while the glue contributes significantly when $c>c^\ast_1$. 

\begin{figure}[h!]
\begin{center}
\includegraphics[width=17cm]{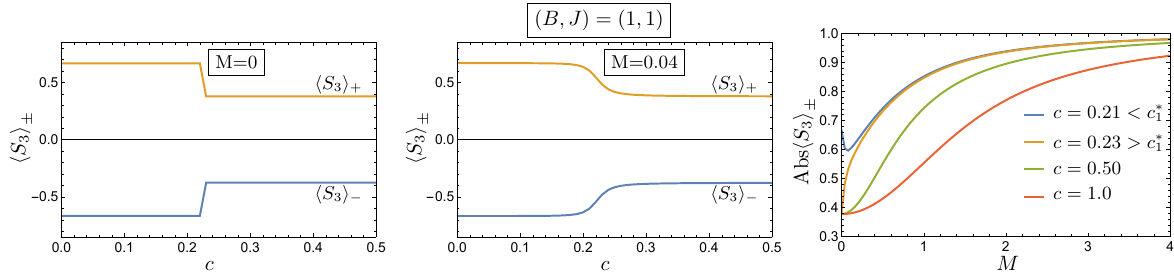}
\vspace*{-0.25cm}
\caption{$\langle S_3 \rangle_{\pm 1}$  as a function of $c$ and $M$ for $(B,J)=(1,1)$.} \label{fig_12}
\end{center}
\end{figure}

The distribution of the spin can be further clarified  by inspecting $\langle S_3\rangle_{J_3}$ (Fig.~\ref{fig_12}).  
We find that when $M=0$, $\langle S_3 \rangle_{\pm 1}$ is discontinuous at $c^\ast_1$:
\begin{eqnarray}
\langle S_3 \rangle_{\pm} = \left\{
\begin{array}{ll}
\pm 0.67 \quad & \text{for } c< c^\ast_1, \\ 
\pm 0.33 \quad & \text{for } c> c^\ast_1.
\end{array}
\right.
\end{eqnarray}
It is unnecessary to plot   $\langle L_3\rangle_{J_3}$ because $\langle L_3\rangle_{J_3}=J_3 - \langle S_3 \rangle_{J_3}$.  


Slightly away from the chiral limit, the discontinuity in  $\langle L^2\rangle$,  $\langle S^2\rangle$ and $\langle S_3 \rangle_{\pm 1}$ smoothens to a crossover  (Fig.\ref{fig_11}-\ref{fig_12}). 
 Thus  in the near-chiral limit, the quark contributes $\sim 67\%$ to the total spin  when $c< c^\ast_1$ and as low as only  $\sim 33\%$ when  $c> c^\ast_1$.

As $M$ increases, the contribution of glue to the total spin decreases, while that of the quark increases. It is not surprising that in the heavy quark limit $M\gg1$,  $\langle L^2\rangle$ becomes tiny, while $\langle S^2\rangle \approx 2$  and $\langle S_3 \rangle_{\pm 1} \approx \pm 1$,  irrespective of the value of $c$.

\subsection{The $(B,J)=(0,1)$ sector} 
\begin{figure}[t!]
\begin{center}
\includegraphics[width=14cm]{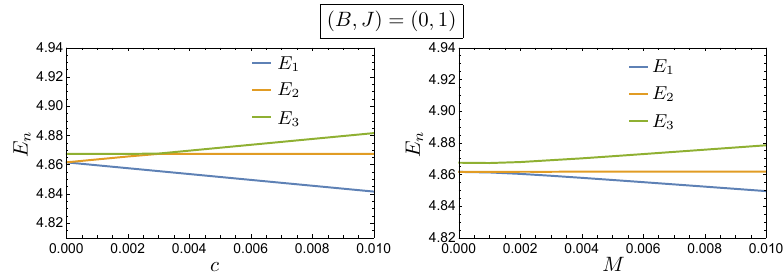}
\vspace*{-0.25cm}
\caption{Lightest three energy levels for $(B,J)=(0,1)$ as a function of $c$ and $M$.} \label{fig_13}
\end{center}
\end{figure}

\begin{figure}[h!]
\begin{center}
\includegraphics[width=17cm]{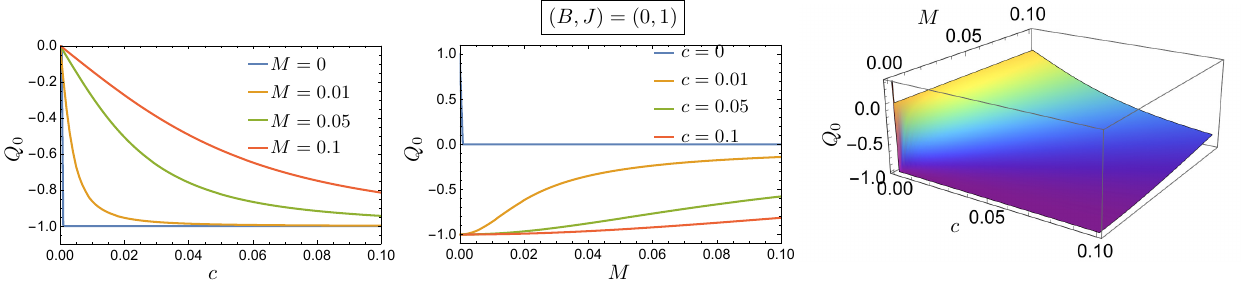}
\vspace*{-0.25cm}
\caption{$Q_0$  as a function of $c$ and $M$  for  $(B,J)=(0,1)$.} \label{fig_14}
\end{center}
\end{figure}

\begin{figure}[t!]
\begin{center}
\includegraphics[width=17cm]{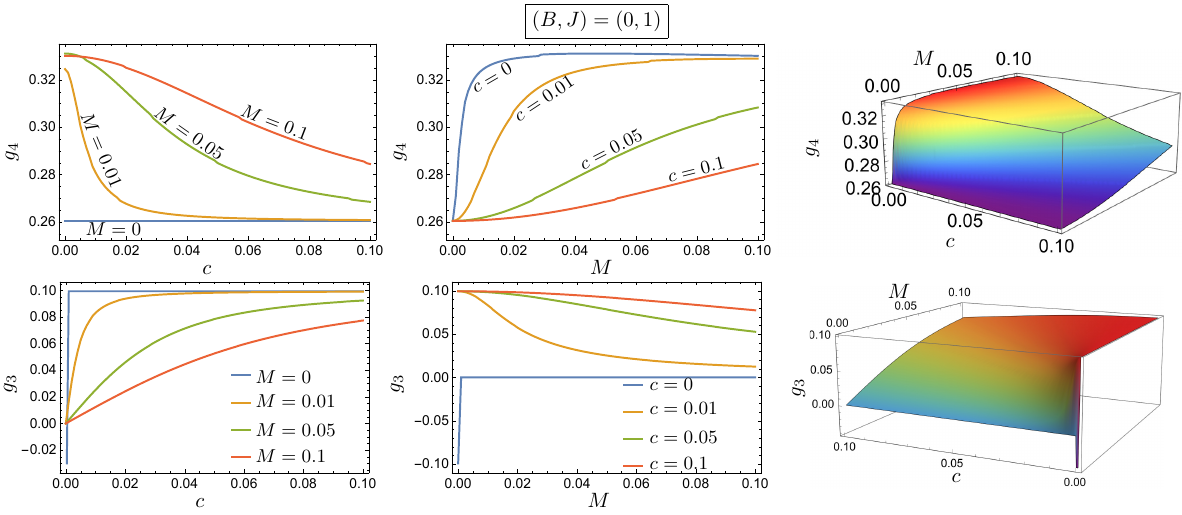}
\vspace*{-0.25cm}
\caption{$g_3$ and $g_4$ as functions of $c$ and $M$ for $(B,J)=(0,1)$.} \label{fig_15}
\end{center}
\end{figure}

 Here too the hadron is a spin triplet.   At the point  $(c, M)=(0,0)$   the ground state is doubly degenerate (see Fig.~\ref{fig_13}). The next excited state at the critical point is almost  degenerate with the (doubly degenerate) ground state, but  we cannot  numerically establish  whether the degeneracy is exact.  At the critical point $(c, M)=(0,0)$, the chiral charge $Q_0$ is discontinuous (see Fig.~\ref{fig_14}). Further, both $g_3$ and $g_4$ display singular behavior at this point, as shown in Fig.~\ref{fig_15}. 

When the quark is massive, the phase transition softens to a smooth crossover.

\begin{figure}[h!]
\begin{center}
\includegraphics[width=17cm]{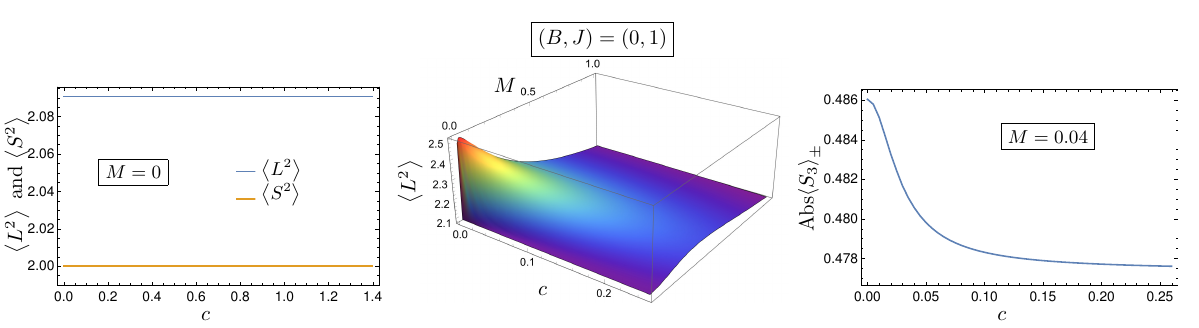}
\vspace*{-0.25cm}
\caption{$\langle L^2 \rangle$, $\langle S^2 \rangle$  and $|\langle S_3\rangle_{\pm}|$ for $(B,J)=(0,1)$ as a function of $c$ and $M$.} \label{fig_16}
\end{center}
\end{figure}

The contribution of quark and glue to total spin can be estimated by computing $\langle L^2 \rangle$, $\langle S^2 \rangle$  and $\langle S_3\rangle_{J_3}$ (see Fig.~\ref{fig_16}). We find that in the chiral limit, $\langle L^2 \rangle$, $\langle S^2 \rangle$  and $\langle S_3\rangle_{J_3}$ remain nearly constant as a function of $c$: 
\begin{eqnarray}
 \langle L^2 \rangle \approx 2.1, \quad  \langle S^2 \rangle\approx 2.0, \quad \langle S_3\rangle_{\pm 1}  \approx \pm 0.5.
\end{eqnarray}



In the heavy quark limit, we expect the quark and the glue to ``decouple''. Surprisingly, we find that both the quark and the glue contribute almost equally to the total spin, indicating an entangled quark-glue state.

\subsection{$(B,J)=(1,0)$ and $(2,0)$ sectors }

In both these sectors, the hadrons are 4-fermion (quark + anti-quark) states and always have $Q_0=0$. In particular,  the 4-fermion state in the $(B,J)=(2,0)$ sector is always the fermion multiplet with $B=2$ (see Table \ref{B2_state}), which is both colorless and spin-0. Consequently, the different hadrons (the ground state and its excitations) of this sector are separable states with different spin-0 color singlet glue states.

The 4-fermion state for the $(B,J)=(1,0)$ has spin-1 color-1 (see Table \ref{B1_states}). Unlike the  $(B,J)=(2,0)$ case, the total state is not separable (i.e. is entangled) and the glue is always in spin-1 color-1 state. 

From the way these states are constituted, it is not difficult to see that  $H_c$, $H_{m}$ and $H_{int}$ vanish identically in these sectors.  These hadrons  are energy eigenstates of pure Yang-Mills theory. Indeed, it is easy to verify that the  hadrons of $(B,J)=(2,0)$ sector have the same energy as the colorless spin-0 glueballs of pure YM. 
 
The $(B,J)=(1,0)$ sector is isospectral with spin-1 color-triplet sector of  pure YM.  Of course, states in the latter are colored and hence not physical. Curiously, adding a quark can neutralize the color  to yield color (and spin)-singlet without any additional energy cost.

Because  $H_m=0=Q_0$ in these sectors, both $M$ and $c$  are irrelevant parameters. Further, for all these hadrons
\begin{eqnarray}
\langle L^2 \rangle =\langle S^2 \rangle=\left\{ \begin{array}{ll}
0 & \quad \text{in } (B,J)=(0,0) \text{ sector} \\ 
1(1+1) & \quad \text{in } (B,J)=(1,0) \text{ sector} 
\end{array}\right.
\end{eqnarray}
There is no level crossing and hence no QPT in these sectors of the theory. Despite its apparently inert nature,  the $(B,J)=(2,0)$ sector  becomes very important when the baryon number chemical potential $\mu$ is switched on, as we will discuss in the next subsection.

\subsection{Baryon number chemical potential}\label{chem_pot_sec}

Let us turn on  the baryon chemical potential by adding $ e_0 \mu  \widehat{B}_3$ (with $\mu \in \mathbb{R}$) to the Hamiltonian (\ref{strong_coupling_Hamiltonian}) (recall that $e_0$ is the energy scale  in  the double scaling limit). This term does not break the $SO(3)_{rot}$ symmetry\footnote{In the field theory counterpart, the term $(\mu Q^\dagger Q)$ preserves rotational symmetry but breaks full Lorentz invariance.}. 
However, because $\widehat{B}_3$ does not commute with $\widehat{B}_1$ and $\widehat{B}_2$, the global  $SU(2)_B$ symmetry is explicitly broken to a $U(1)_B$.  The effect of this symmetry breaking is reflected in the spectrum of the hadrons: the degeneracy between states  with different $B_3$ in any hadron multiplet is lifted.  For a hadron with any baryon charge, the energy $E(\mu) $ of the state is 
\begin{eqnarray}
E(\mu) = E(\mu=0) + \mu B_3.
\end{eqnarray}

\begin{figure}[h!]
\begin{center}
\includegraphics[width=14cm]{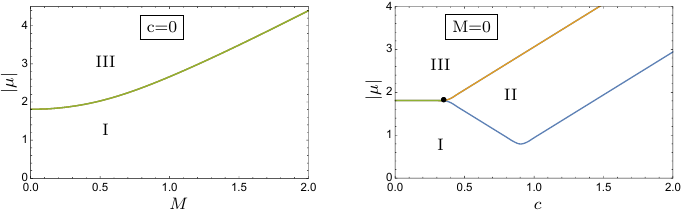}
\vspace*{-0.5cm}
\caption{Phase diagram in presence of chemical potential when $c=0$ (left) and when $M=0$ (right). The green, blue and orangle curves are  the coexistence lines between the phases. The black dot represents the triple point where all three phases  coexist.  } \label{fig_phase_diag}
\end{center}
\end{figure}

The singlet states with $B=0$ are unaffected by $\mu$. But for the sectors with $B=1$ and $2$, the lightest energy eigenstate is no longer the meson. With positive/negative $\mu$, the lightest state is an anti-diquark/diquark for $B=1$ sectors, and an anti-tetraquark/teteraquark for the $B=2$ sector.

Since the lightest state of $(B,J)=(1,1)$  is lighter than that of $(B,J)=(1,0)$ (see Fig.~\ref{Fig_1}),   the lightest diquark/anti-diquark is always a spin-1 state for any $\mu$. 

However,  if $|\mu|$ is sufficiently large, the lightest meson of the $(B,J)=(0,0)$ sector can become heavier than the lightest diquark/anti-diquark from the $(B,J)=(1,1)$ sector, or even the tetraquark/anti-tetraquark from the $(B,J)=(2,0)$ sector, depending on the values of $c$ and $M$. When this happens, the  ground state of the theory belongs to $(B,J)=(1,1)$ or $(2,0)$ sector. Thus with $\mu$ turned on, there can be three distinct phases of the theory (see Fig.~\ref{fig_phase_diag}): 
\begin{enumerate}[i)]
\item Phase I -- the ground state is a meson from $(B,J)=(0,0)$ sector. 
\item Phase II -- the ground state is a diquark/anti-diquark from the $(B,J)=(1,1)$ sector. 
\item Phase III -- the ground state is a tetraquark/anti-tetraquark from the $(B,J)=(2,0)$ sector. 
\end{enumerate}
If we tune $|\mu|$ keeping $c$ and $M$ fixed,  the lowest energy levels of $(B,J)=(0,0), (2,0)$ and/or $(1,1)$ can cross, yielding first order QPTs between the phases.

When $c=0$, tuning $\mu$ and $M$ can induce a transition between phases I and III, and phase II  is absent (see Fig.~\ref{fig_phase_diag}).    In contrast, when $c\neq 0$, all three phases can exist and there is a triple point. 

Interestingly, when the system is in phase II, the ground state is a spin-1 triplet and hence breaks the rotational invariance. 
 The Fig.~\ref{fig_phase_diag} clearly shows that in the chiral limit, phase II emerges only for $c\gtrsim 0.4 >c_1^\ast$. For  $c>c_1^\ast $, the ground state has $Q_0=-1$,  and hence is a two-fermion state  (see  Fig.~\ref{fig_9}). Thus the ground state in the chiral limit has a pair of quarks/anti-quarks (for $\mu$ negative/positive).  Further, the fermionic spin of the quark/anti-quark pair in the triplet state is non-zero: $\langle S_3\rangle_{\pm 1} = \pm 0.33$ (as shown in Fig.~\ref{fig_12}). 
 
 This phase is reminiscent of the LOFF phase as the ground state breaks $SO(3)_{rot}$. This result also conforms with the findings in \cite{Fukushima:2007bj, Andersen:2010vu, Kanazawa:2020ktn}, although the model there is quite different.

\section{Summary and Discussion}

We have presented an exhaustive and complete discussion of the phase structure of matrix-QCD$_{2,1}$ in the strong coupling regime. The robustness of the quantum phase transition is established by studying, in addition to the ground state energy, several other observables like chiral charge and the higher Binder cumulants. When the baryon chemical potential $\mu$ is sufficiently large, there is a LOFF-like phase with a spin-1 ground state. 

In the $(B,J)=(0,0)$ sector, the chiral phase with $c<c_0^\ast$ has an entangled quark-glue ground state, whereas it is a scalar glueball (i.e. quark-glue interaction energy is zero) for $c>c_0^\ast$.  Interestingly, in the entangled ground state for $c<c_0^\ast$, the third and fourth Binder cumulants indicate that the gauge configurations are localized near a  special corner of the configuration space where the matrix degrees of freedom corresponds to only one non-zero singular value.

In the sectors that we have studied,  the glue carries a significant share of the hadron's spin, especially when the quark is very light. In particular, we have found that in the chiral ground state of $(B,J)=(1,1)$ sector, the contribution of the quark to the spin of the state can be as low as 33\%.



It would be interesting to extend the two-color QCD to more flavors or to situations with non-trivial supersymmetry. Indeed, the supersymmetric version of the matrix model \cite{ErrastiDiez:2020iyk} is currently under numerical investigation and results will be reported separately.



Finally,  our numerical techniques can be extended to real world QCD to investigate its phase structure, exotic states like tetraquarks and pentaquarks, and the proton spin question.  The matrix model provides a simplified framework to quantitatively address this set of questions.

\mbox{}\\
\textbf{Acknowledgements:} It is our great pleasure to thank Denjoe O’Connor and V. Parameswaran Nair for discussions and suggestions. NA would like to acknowledge financial support (grant no. SP-111) from IIT Bhubaneswar to acquire  computational resources.

\appendix
\section*{Appendices} 
\renewcommand\thefigure{\thesection.\arabic{figure}} 
\setcounter{figure}{0}    
\section{Convergence of the energy eigenvalues} \label{app:convergence} 
\begin{figure}[h!]
\begin{center}
\includegraphics[width=8cm]{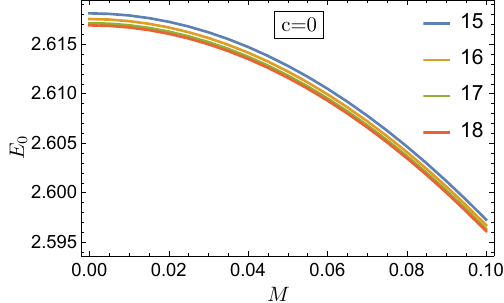}
\vspace*{-0.25cm}
\caption{The ground state energy for $(B,J)=(1,1)$ sector vs $M$ at $c=0$ with $N_{max}=15,16,17,18$. }\label{convergence_fig} 
\end{center} 
\end{figure}

We have performed the numerical analysis with $10\leq N_{max} \leq 18$. As $N_{max}$ increases, the energy eigenvalues progressively converge. In fact there there is hardly any difference between $ N_{max} =17$ and $18$, as can be seen in Fig.~\ref{convergence_fig}. For instance, at $M=0.05$, the ratio 
\begin{eqnarray}
\frac{E_0(N_{max}=17) -E_0(N_{max}=18)}{E_0(N_{max}=17)} =  \frac{2.6122 - 2.6120}{2.6122} = 8 \times 10^{-5}
\end{eqnarray}
The convergence is similarly excellent for the ground state energy (and expectation values of other observables) in other sectors as well.  For brevity, we only show the data for the ground state energy in $(B,J)=(1,1)$ sector.


\section{Quark states} \label{app:states} 
The 8-fermion state is defined as
\begin{eqnarray}
|8_{F}\rangle&\equiv& b^{\dagger}_{11}\,b^{\dagger}_{12}\,b^{\dagger}_{21}\,b^{\dagger}_{22}\,d^{\dagger}_{11}\,d^{\dagger}_{12}\,d^{\dagger}_{21}\,d^{\dagger}_{22} |0_{F}\rangle
\end{eqnarray} 

We can define the two-fermion mesonic, di-quark and anti-di-quark operators as
\begin{eqnarray}
\mathcal{S}_{\mu \nu}&\equiv& b^{\dagger}_{\alpha A}\,\sigma^{\mu}_{\alpha \beta}\,T^{\nu}_{A B}\,d^{\dagger}_{\beta B}\,, 
\\
\Gamma_{\mu \nu}&\equiv& b^{\dagger}_{\alpha A}\,(\sigma^{\mu}\,\sigma^{2})_{\alpha \beta}\,(T^{\nu}T^{2})_{A B}\,b^{\dagger}_{\beta B}, \\
\Lambda_{\mu \nu} &\equiv& d^{\dagger}_{\alpha A}\,(\sigma^{2}\,\sigma^{\mu})_{\alpha \beta}\,(T^{2}T^{\nu})_{A B}\,d^{\dagger}_{\beta B}
\end{eqnarray}
where $\mu,\nu=0,1,2,3$ with $T^0=\frac{1}{2} I_2$ and $\sigma^0= I_2$

The fermionic states are labelled by their spin and color: (s,c) denote a spin-s state which transform in the (2c+1)-dimensional representation of color $SU(2)$. 

Fermion states with with baryon charge $B$ are given in Table \ref{B0_states}- \ref{B2_state}. 
\begin{center}
\begin{table}[H]
\begin{center}
{\small \begin{tabular}{|c|c|c|c|} \hline 
\,\,  Spin-0 Color-0\,\,  & \,\,Spin-1 Color-1 \,\, & Spin-2 Color-0 & Spin-0 Color-2 \\ 
 (0,0) & (1,1) & (2,0) & (0,2) \\ \hline 
$\qquad\ket{0_{F}}\qquad$  & &&\\
$\mathcal{S}_{00}\ket{0_{F}}$ & $\mathcal{S}_{ia}\ket{0_{F}}$ && \\
$\sqrt{\frac{2}{3}}\,\mathcal{S}_{00}\mathcal{S}_{00}\ket{0_{F}}$ & $\sqrt{2}\,\mathcal{S}_{ia}\mathcal{S}_{00}\ket{0_{F}}$ & $\frac{\sqrt{2}}{3}\,\Big(\mathcal{S}_{ic}\mathcal{S}_{jc}-\frac{\delta_{ij}}{3}\mathcal{S}_{kc}\mathcal{S}_{kc}\Big)\ket{0_{F}}$ & $\frac{\sqrt{2}}{3}\,\Big(\mathcal{S}_{ka}\mathcal{S}_{kb}-\frac{\delta_{ab}}{3}\mathcal{S}_{kc}\mathcal{S}_{kc}\Big)\ket{0_{F}}$  \\
$\mathcal{S}^{\dagger}_{00}\ket{8_{F}}$ &  $\mathcal{S}^{\dagger}_{ia}\ket{8_{F}}$ & &   \\
$\qquad\ket{8_{F}}\quad$ & && \\ \hline 
\end{tabular}}
\end{center}
\caption{Fermion states with Baryon charge $B=0$, Note that $B_3=0$ for all the above states. } \label{B0_states} 
\end{table} 
\begin{table}[H]
\begin{center}
{\small  \begin{tabular}{ | *{3}{c |} *{3}{c |}}
        \hline
       
           \multirow[c]{1}{*}{  }
        & $B_3=1$ & $B_3=0$ & $B_3=-1$ \\ \hline 
        \multirow[c]{3}{*}{ Spin-1 Color-0: (1,0)}
        & $\sqrt{2}\,\Gamma_{i0}\ket{0_{F}}$ & $\mathcal{S}_{i0}\ket{0_{F}}$ & $\sqrt{2}\,\Lambda_{i0}\ket{0_{F}}$ \\
&$2\,\Gamma_{i0}\mathcal{S}_{00}\ket{0_{F}}$ & $\sqrt{2}\,\mathcal{S}_{i0}\mathcal{S}_{00}\ket{0_{F}}$ & $2\,\Lambda_{i0}\mathcal{S}_{00}\ket{0_{F}}$ \\
& $\sqrt{2}\,\Lambda^{\dagger}_{i0}\ket{8_{F}}$ & $\mathcal{S}_{i0}^{\dagger}\ket{8_{F}}$ & $\sqrt{2}\,\Gamma_{i0}^{\dagger}\ket{8_{F}}$
             \\ \hline
             
   \multirow[c]{1}{*}{ Spin-1 Color-1: (1,1)}           & $\epsilon_{ijk}\,\mathcal{S}_{ja}\Gamma_{k0}\ket{0_{F}}$  & $\frac{1}{\sqrt{2}}\,\epsilon_{ijk}\,\mathcal{S}_{ja}\mathcal{S}_{k0}\ket{0_{F}}$ & $\epsilon_{ijk}\,\mathcal{S}_{ja}\Lambda_{k0}\ket{0_{F}}$ \\ \hline 
   \multirow[c]{3}{*}{Spin-0 Color-1:  (0,1)} & $\sqrt{2}\,\Gamma_{0a}\ket{0_{F}}$ & $\mathcal{S}_{0a}\ket{0_{F}}$ & $\sqrt{2}\,\Lambda_{0a}\ket{0_{F}}$ \\
   & $2\,\Gamma_{0a}\mathcal{S}_{00}\ket{0_{F}}$ & $\sqrt{2}\,\mathcal{S}_{0a}\mathcal{S}_{00}\ket{0_{F}}$ & $2\,\Lambda_{0a}\mathcal{S}_{00}\ket{0_{F}}$ \\
   & $\sqrt{2}\,\Lambda^{\dagger}_{0a}\ket{8_{F}}$ & $\mathcal{S}_{0a}^{\dagger}\ket{8_{F}}$ & $\sqrt{2}\,\Gamma_{0a}^{\dagger}\ket{8_{F}}$ \\ \hline 
    \end{tabular}}
    \end{center}
    \caption{Fermion states with Baryon charge $B=1$.} \label{B1_states}
    \end{table}

\begin{table}[H]
\begin{center}
{\small \begin{tabular}{|c|c|c|c|} \hline 
& Spin-0 Color-0: (0,0)   \\ \hline 
$B_3=2$ & $\quad\frac{2}{3}\,\Gamma_{0a}\Gamma_{0a}\ket{0_{F}}\qquad$ \\
$B_3=1$ & $\qquad\frac{2}{3}\,\mathcal{S}_{0a}\Gamma_{0a}\ket{0_{F}}\quad$ \\
$B_3=0$  &  $\frac{1}{2\sqrt{6}}\,(\mathcal{S}_{i0}\mathcal{S}_{i0}-\mathcal{S}_{0a}\mathcal{S}_{0a})\ket{0_{F}}$ \\
$B_3=-1$ &  $\qquad\frac{2}{3}\,\mathcal{S}_{0a}\Lambda_{0a}\ket{0_{F}}\quad$ \\
$B_3=-2$ & $\qquad\frac{2}{3}\,\Lambda_{0a}\Lambda_{0a}\ket{0_{F}}\quad$\\ \hline 
\end{tabular}} 
\caption{Fermion states with Baryon charge $B=2$.}\label{B2_state}
\end{center}
\end{table}

\end{center}

\section{The spin-0 and spin-1 colorless states}

\begin{table}[H]
The glue  states (up to spin-2 and color-2) in $SU(2)$ matrix model are of the following (spin, color) type:  \\ 
\begin{center}
{\small  \begin{tabular}{ | *{3}{c |} *{3}{c |}}
        \hline
       
         \multirow[c]{1}{*}{ \bf Spin  } & {\bf Color }& {\bf Glue state} \\ \hline
        \multirow[c]{2}{*}{ Spin-0  }
        & Color-0 &  (0,0)  \\
        &Color-2 &  (0,2)   \\\hline 
          \multirow[c]{2}{*}{ Spin-1  }
        & Color-1 &  (1,1)  \\
        &Color-2 &  (1,2)   \\\hline
        \multirow[c]{3}{*}{ Spin-2  }
        & Color-0 &  (2,0)  \\
        & Color-1 &  (2,1)  \\
        &Color-2 &  (2,2)   \\\hline
    \end{tabular}}
\end{center}
\end{table} 
Note that in QCD$_{2,1}$ there are \textit{no} glue states of the type $(0,1)$ and $(1,0)$. 

%

\begin{table}[H]
The spin-0 and spin-1 colorless states in the  multiplets with different values of $B$ are composed of the following states: 
\begin{center}
{\small  \begin{tabular}{ | *{3}{c |} *{3}{c |}}
        \hline
       
           \multirow[c]{1}{*}{  }
     {\bf   Sectors }& \quad   {\bf  Quark states}  \quad & \quad  {\bf   Glue states} \quad   \\ \hline 
        \multirow[c]{4}{*}{ $B=0$ \quad \quad $J=0$ }
        & (0,0) &  (0,0)  \\
        & (1,1) &  (1,1)  \\ 
        & (2,0) &  (2,0)  \\
        & (0,2) &  (0,2)  \\\hline 
          \multirow[c]{4}{*}{ $B=0$ \quad \quad $J=1$ }
        & (0,0) &  Not possible  \\
        & (1,1) &  (1,1) and (2,1)  \\ 
        & (2,0) &  (2,0)  \\
        & (0,2) &  (1,2)  \\\hline 

       
        \multirow[c]{3}{*}{ $B=1$ \quad \quad $J=0$ }
        & (1,0) &  Not possible   \\
        & (1,1) &  (1,1)  \\ 
        & (0,1) &  Not possible \\\hline 
          \multirow[c]{3}{*}{ $B=1$ \quad \quad $J=1$ }
        & (1,0) &  (0,0) and (2,0)  \\
        & (1,1) &  (1,1) and (2,1)  \\ 
        & (0,1) &  (1,1)  \\ \hline 
          \multirow[c]{1}{*}{ $B=2$ \quad \quad $J=0$ }
        & (0,0) &  (0,0)   \\ \hline  
    \end{tabular}}
\end{center} 
\end{table} 

\end{document}